\documentclass[conference]{IEEEtran}

\usepackage[utf8]{inputenc} 
\usepackage{indentfirst}
\usepackage{mathptmx}
\usepackage{graphicx}
\usepackage[labelfont=bf]{caption}

\title{Classification of COVID-19 in CT Scans using Multi-Source Transfer Learning}

\author{
  Alejandro R. Martinez \\
  Dartmouth College\\
  \texttt{alejandro.r.martinez.20@dartmouth.edu} \\

}

\begin{document}

\date{}
\maketitle
\begin{abstract}
Since December of 2019, novel coronavirus disease COVID-19 has spread around the world infecting millions of people and upending the global economy. One of the driving reasons behind its high rate of infection is due to the unreliability and lack of RT-PCR testing. At times the turnaround results span as long as a couple of days, only to yield a roughly 70\% sensitivity rate. As an alternative, recent research has investigated the use of Computer Vision with Convolutional Neural Networks (CNNs) for the classification of COVID-19 from CT scans. Due to an inherent lack of available COVID-19 CT data, these research efforts have been forced to leverage the use of Transfer Learning. This commonly employed Deep Learning technique has shown to improve model performance on tasks with relatively small amounts of data, as long as the Source feature space somewhat resembles the Target feature space. Unfortunately, a lack of similarity is often encountered in the classification of medical images as publicly available Source datasets usually lack the visual features found in medical images. In this study, we propose the use of Multi-Source Transfer Learning (MSTL) to improve upon traditional Transfer Learning for the classification of COVID-19 from CT scans. With our multi-source fine-tuning approach, our models outperformed baseline models fine-tuned with ImageNet. We additionally, propose an unsupervised label creation process, which enhances the performance of our Deep Residual Networks. Our best performing model was able to achieve an accuracy of 0.893 and a Recall score of 0.897, outperforming its baseline Recall score by 9.3\%.
\end{abstract}
\vspace{2.5mm}


\vspace{2.5mm}

\section{Introduction \& Related Work}
On March 11th 2020, the World Health Organization (WHO) proclaimed the novel coronavirus disease COVID-19 a global pandemic. Originating in the Hubei Province of China in late 2019, COVID-19 has spread across 185 countries, infecting over 30 million people and causing nearly 1 million deaths \cite{who,jhu}. One of the main reasons for its unprecedented growth is due to the unreliability and lack of testing \cite{correlat}. 

The most widely employed test kits are reverse transcription polymerase chain reaction (RT-PCR) assays, which check for the detection of nucleic acid from SARS-CoV-2 in respiratory specimens \cite{fda}. While RT-PCR Assays are commonly used, they are reported to yield poor sensitivity in early stages of infection and require a lengthy processing time \cite{correlat, fang2020sensitivity}. In addition to these issues, RT-PCR Assays face severely limited supply, causing many symptomatic people to be left untested \cite{resource}.

In light of these constraints, Computed Tomography (CT) imaging has been explored as a possible alternative diagnostic tool for COVID-19 \cite{correlat, fang2020sensitivity, hope2020chest}. Prominent features of the virus, such as bilateral ground-glass opacities, have been identified in the chest CT scans of patients with COVID-19. These visual features have a potential to act as regions of interest in the detection of the virus \cite{correlat, fang2020sensitivity}. CT imaging also produces much faster results in comparison to RT-PCR Assays and is widely available with roughly 6,000-7,000 scanners present in the United States \cite{industry}.

The use of CT imaging as a diagnostic tool would require Deep Learning and Computer Vision technologies. These computational tools have been successfully applied to CT and other medical imaging classification tasks with low rates of error \cite{suzuki2017overview,yamashita2018convolutional,song2017using}. The most commonly applied algorithm for image classification is the Convolutional Neural Network (CNN). Researchers have used CNNs for classification of lung diseases in chest CT with radiologist level accuracy \cite{rajpurkar2017chexnet}. Recently, CNNs have been applied to COVID-19 chest CT classification and have shown promising results. He et al (2020) and Xu et al (2020) have both explored the use of CNNs for distinguishing COVID-19 from other types of pneumonia or normal chest CTs and managed to achieve overall accuracies of 86.0\% and 86.7\%, respectively \cite{he2020sample,butt2020deep}. While recent research has seemed hopeful, there still remain limitations. 

With all Deep Learning problems, the amount of collected data largely determines the success of the system. As COVID-19 is a novel virus, there is an inherent lack of available datasets to construct a robust Deep Learning classifier capable of producing reliable results. To compensate for this issue, both He et al (2020) and Xu et al (2020) exploit the use of transfer learning: a method by which a network is pretrained on a large Source task and then retrained on a smaller Target task. Transfer learning provides a network with a deep understanding of generic features from a Source dataset, so it does not require much data to learn the idiosyncrasies of the Target dataset \cite{pan2009survey}. The problem with applying this method to medical imaging, however, is that the Source dataset the network is trained on (i.e. ImageNet) usually contains very dissimilar feature spaces to those in the Target dataset \cite{yamashita2018convolutional}. This leaves the network with sub par performance, compared to what it could achieve if pretrained on an additional dataset of medical images.

In this study we aim to provide a highly sensitive classification model for the detection of COVID-19 by exploiting a Multi-Source Transfer Learning (MSTL) process to distinguish COVID-19 from normal CT scans. 

We start with the collection of multiple Deep CNNs pretrained on ImageNet, provided by TensorFlow, Google's open source Machine Learning Library \cite{tf}. We then collect two datasets: the first is comprised of 22,238 lung CT slices from the SPIE-AAPM Lung CT Challenge provided by the Cancer Imaging Archive; the second is comprised of 349 COVID-19 and 397 normal CT scans provided by the UCSD Department of Engineering \cite{lungx, zhao2020COVID-CT-Dataset}. The first dataset is used to teach our pretrained ImageNet models to extract relevant features from chest CT scans, and the second dataset is used to further fine-tune the models on distinguishing COVID-19 from normal chest CT scans. 
\vspace{2.5mm}

\section{Methods}
In this section we describe our Multi-Source Transfer Learning (MSTL) approach for the classification of COVID-19 from normal chest CT scans. Our methodology begins with the collection of multiple chest CT datasets, followed by data preprocessing, model selection, and ultimately MSTL.

\subsection{Data description}

Our study utilizes three separate datasets as part of our multi-source fine-tuning paradigm: a source, a transition, and a target. 

\textbf{Source dataset:} Being that our selected models are pretrained on an ImageNet subset known as the ImageNet Large Scale Visual Recognition Challenge (ILSVRC) dataset, this will serve as our Source dataset. The ILSVRC dataset comprises of 1.2 million images spanning 1,000 unique classes. Each class in the dataset corresponds to a distinct synonym set, or a synset, defined by WordNet, a large lexical database that retains a semantic hierarchy between concepts through the construction of synsets \cite{imagenet}. This structure ensures that ILSVRC classes hold unique feature representations, making the dataset conducive to generalization. All images are labeled by human annotators via Amazon's Mechanical Turk, a crowd sourcing marketplace \cite{imagenet,amazon}.

\textbf{Transition dataset:} Our Transition dataset comes from the 2015 SPIE-AAPM-NCI Lung Nodule Classification Challenge, made available through The Cancer Imaging Archive (TCIA) and sponsored by the SPIE, NCI/NIH, AAPM and The University of Chicago. The dataset contains 22,489 CT scan slices from 70 patients (28 males, 42 females: median age: 61 years), containing 42 benign and 41 malignant lung nodules in total. All scans were acquired on Philips Brilliance 16, 16P, and 64 scanners, and stored as 3-dimensional DICOM files with resolution of 512x512 per slice. All protected health information was removed from DICOM headers \cite{lungx}.

\textbf{Target dataset:} Our Target dataset is collected from an open access COVID-19 CT image repository provided by the UCSD Department of Engineering \cite{zhao2020COVID-CT-Dataset}. The dataset contains 349 COVID-19 and 397 nonCOVID-19 or normal CT scans. The normal CT scans were collected from MedPix, an open-access online database of medical images. The COVID-19 scans were manually selected from 760 preprints on COVID-19 from medRxiv and bioRxiv, published from January 19th to March 25th. Of the scans collected, 137 contain gender information and 169 contain age information. From the available metadata, the mean age of patients is calculated to be roughly 45 years old and the gender distribution is 86 males to 51 females. Most cases were reported to be from East Asia, with an overwhelming majority from Wuhan, China. It should be noted that the creators of the dataset claim the quality of CT images is well-preserved.

\begin{figure}[h]
\centering
\includegraphics[scale=0.3]{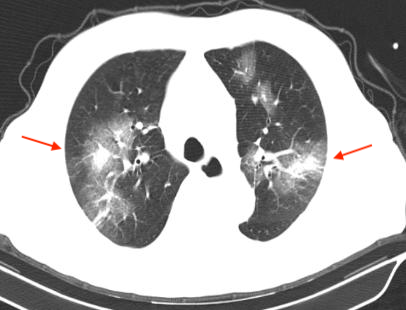}
\label{fig:Fig1}
\caption{Regions of interest exhibiting key identifiers of COVID-19}
\end{figure}

\subsection{Data preprocessing}

The purpose of the Transition step is to improve the learning transfer from the Source task to Target task. It does this by teaching our network to extract low to mid-level features that are more prevalent in our Target feature space than in our Source feature space. Therefore, it is essential that the Transition dataset is thoroughly filtered of images that may be too dissimilar from images in our Target dataset, so the model only learns relevant features during the Transition step.

Scans from the Target dataset contain only 2-dimensional slices, focusing on areas of the lungs displaying distinguishable COVID-19 symptoms. This region of interest captures a clear view of lung lobes that would exhibit key identifiers of COVID-19 such as bilateral ground glass opacities, as depicted in Fig.1. Because our Transition dataset comprises of 3-dimensional CT scans, containing many axial slices per scan, we excluded any slices that did not display the same regions of interest as shown in the Target slices. We then exported the resulting 10,176 Transition slices as JPG files to process them as image arrays in our pretrained networks. We conducted all preprocessing on Transition data with Horos.

To diminish the variability in image processing, we ensured that each input image was reshaped to a standard dimensions of (224,224,3).

\begin{figure}[h]
\centering
\includegraphics[height=1.8cm,width=\linewidth]{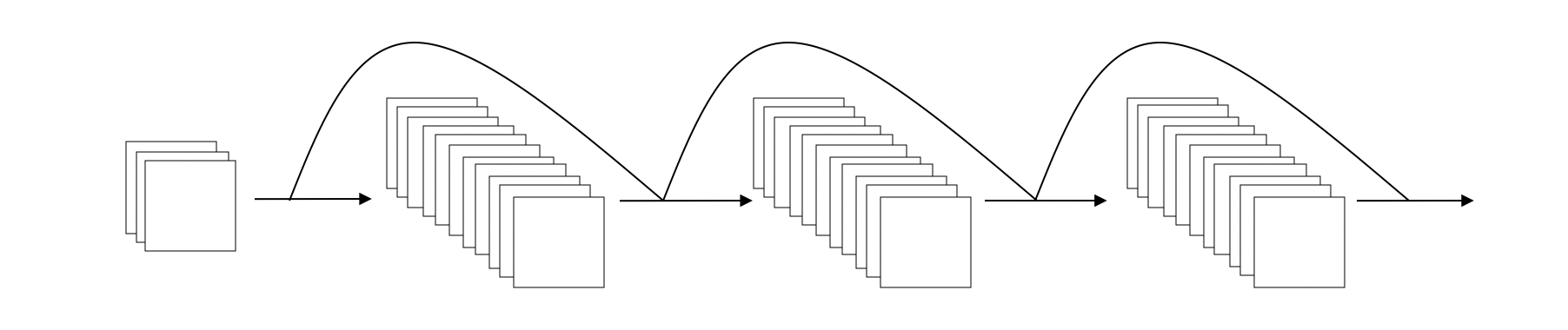}
\includegraphics[height=3cm,width=\linewidth]{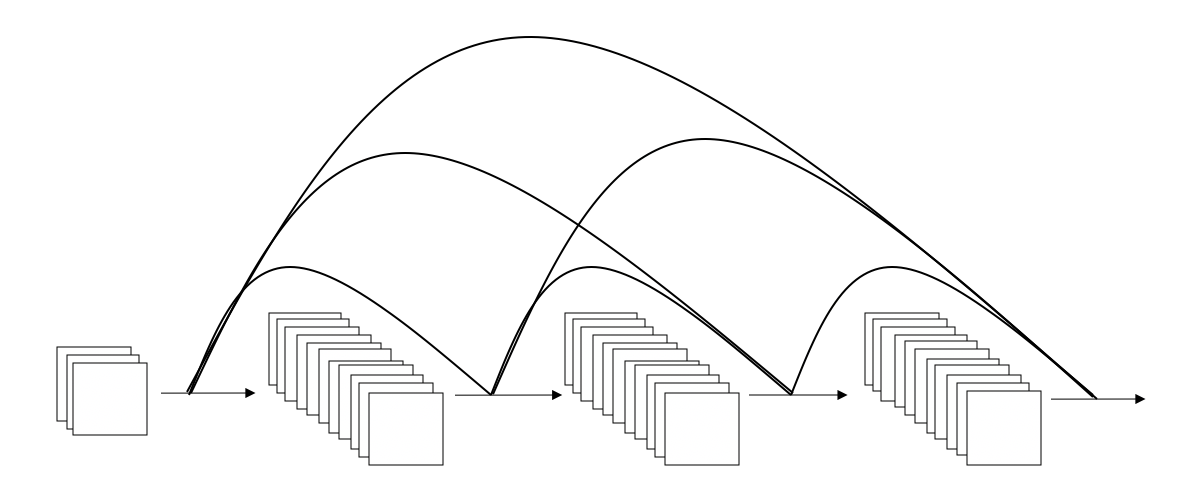}
\label{fig:Fig2}
\caption{ResNet (top) \& DenseNet block (bottom) Architectures}
\end{figure}

\subsection{Model Selection}

We selected four pretrained models from Tensorflow's Keras API: ResNet50V2, ResNet101V2, DenseNet121, and DenseNet169 \cite{resnet,densenet}. See Figure 2 for a comparison of architectures.

\textbf{ResNet:} The ResNet is a Deep Residual Network, a type of CNN often employed in the field of computer vision since gaining recognition from wining the 2015 ILSVRC \cite{resnet}. Deep Residual Networks act similarly to deep CNNs except they implement a residual connection between any given layer and its following layer's output. This entails that any given layer will receive feature maps as input from its two preceding layers. The residual connections improve upon regular neural networks in two ways: they mitigate the vanishing gradient problem by allowing the use of an alternative paths for gradient flow, and they allow the model to learn referenced functions which ensures deeper layers will perform either better or at least as good as shallower layers \cite{resnet}.

\textbf{DenseNet:} The DenseNet is very similar to the ResNet, however, instead of retaining a single residual connection between any given layer and its following layer's output, any given layer in a DenseNet retains a residual connection from each of its preceding layers. This entails that any n\textsuperscript{th} layer will take feature maps as input from n-1 preceding layers. The Dense connectivity of this network is then compacted by grouping dense layers into \emph{dense blocks} and introducing transition layers, which apply convolutions and pooling operations to the dense blocks' output feature maps, reducing the depth of these feature maps by a compression factor of $\theta$ \cite{densenet}. This compression process allows DenseNets to hold less parameters than ResNets, as depicted in Table 1.

\begin{table}[h]
\centering
\includegraphics[scale=0.45]{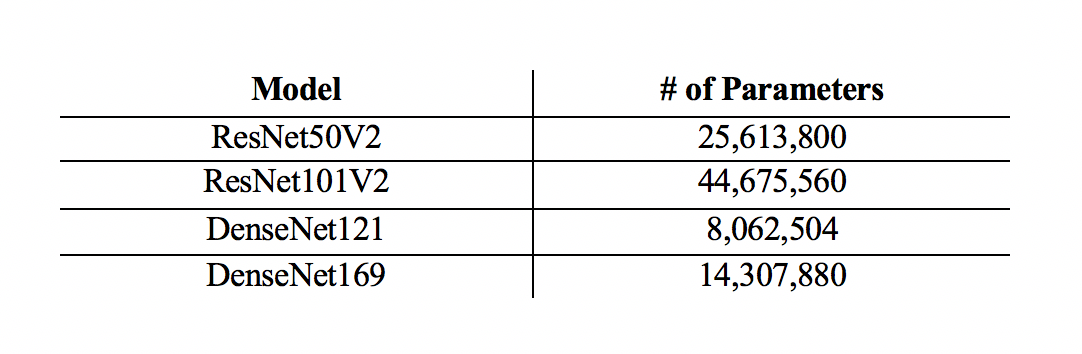}
\label{table:Table1}
\caption{Parameters of each model}
\end{table}

\subsection{Multi-Source Transfer Learning}

As illustrated in Figure 3, our Multi-Source Transfer Learning process involves a three step process:
\begin{enumerate}
   
    \item\textbf{Source step:} a randomly initialized model learns a Source task T\textsubscript{s} on a Source domain D\textsubscript{s}
    \item\textbf{Transition step:} the model is then fine-tuned on a Transition task T\textsubscript{t} on a Transition domain D\textsubscript{t}
    \item\textbf{Target step:} the model is further fine-tuned on a Target task T\textsubscript{targ} on a Target domain D\textsubscript{targ} 
\end{enumerate}

\begin{figure}[h]
\centering
\includegraphics[scale=0.4]{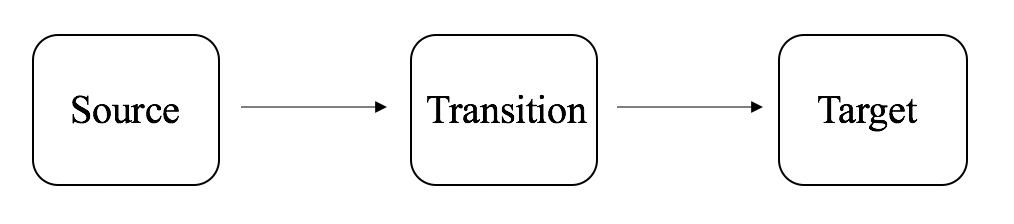}
\label{fig:Fig3}
\caption{Multi-Source Transfer Learning paradigm}
\end{figure}

In theory, the Transition step can be a sequence of steps, however, in this study we limit the total number of Transition steps to 1 for computational simplicity. The Multi-Source process allows for increasingly positive transfer of knowledge at each step assuming that at each step, \emph{i}, the domain D\textsubscript{i} is a better approximation of Target domain D\textsubscript{targ} than its preceding step, \emph{i}-1.

Therefore: 
\begin{equation}
|D\textsubscript{i-1} - D\textsubscript{targ}| > |D\textsubscript{i}-D\textsubscript{targ}| \quad
\end{equation}

,where the difference in similarity between \emph{a} and \emph{b} is represented as: \begin{equation}|\emph{a} - \emph{b}|\end{equation}

\textbf{Source step:} As indicated in section 2.3, our loaded models are pretrained on the ILSVRC dataset. This indicates that our models have already learned our Source task T\textsubscript{s} on our Source domain D\textsubscript{s}.

\textbf{Transition step:} For our Transition step we decided to explore the use of two alternate Transition tasks via different labeling strategies: Soft Labeling and Hard Labeling.
\begin{itemize}
    \item \emph{Hard Labeling:} hard labels can be though of as ground truth labels. In other words, for the Hard Labeling strategy we utilize the original labels that were provided to us in the Transition dataset: malignant (1) or benign (0). This binary labeling strategy makes our Hard Labeling Transition task T\textsubscript{t-hard} a binary classification task.
    
    \item \emph{Soft Labeling:} soft labels can be thought of as a sort of unsupervised labeling strategy that can be applied to unlabeled data. Previous attempts at soft labeling employ the use of an auxiliary task, which teaches a model how to learn unlabeled data when knowledge of the domain is known \cite{pan2009survey}. Alternatively, we explore a soft labeling strategy in which knowledge of the domain is either not known or ignored. 

\begin{figure}[h]
\centering
\includegraphics[scale=0.38,width=\linewidth]{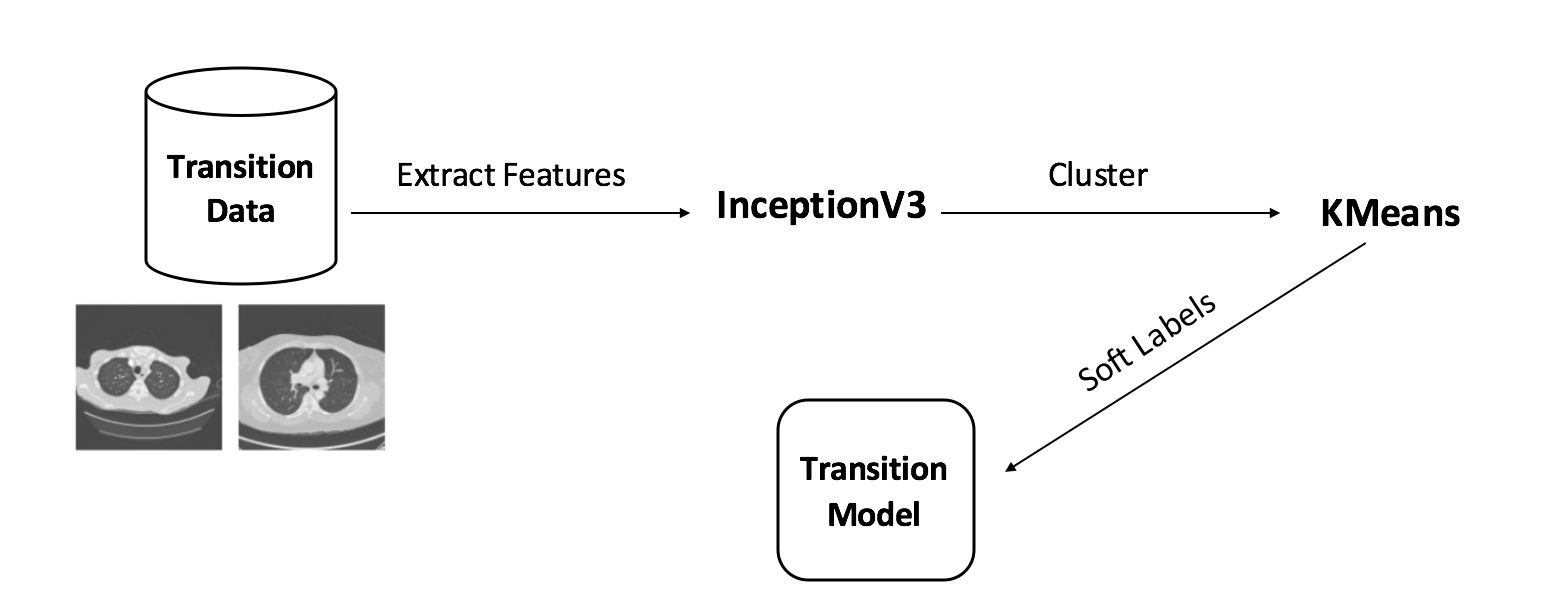}
\label{fig:Fig4}
\caption{Unsupervised creation of Soft Labels}
\end{figure}

    As shown in Figure 4, our soft labeling strategy begins by first feeding the Transition data through a pretrained InceptionV3 convolutional base, provided by Tensorflow \cite{inception}. We utilize the InceptionV3 base as a trained feature extractor, which converts our original input images of dimensions (224,224,3) into feature maps of dimensions (5, 5, 2048). The resulting feature maps are then flattened and fed into a KMeans clustering algorithm, provided by scikit-learn, which clusters the data into 16 distinct feature groups \cite{sklearn}. We then apply corresponding labels to each cluster resulting in 16 labels: making the Transition task T\textsubscript{t-soft} a 16-class classification task. To select the number of clusters (k), we performed a grid search with cross validation of 10, evaluating k values of \{2,4,8,16\}.
\end{itemize}

After acquiring our hard and soft labels, we configured our models to fit the Transition tasks. Because our pretrained models are loaded as convolutional bases we added randomly initialized pooling and fully connected layers, appropriate to each model. For the ResNet models we appended a 2-dimensional average pooling layer, a flatten layer, and a dense layer of 1000 nodes. For the DenseNet models we appended a 2-dimensional global average pooling layer and a dense layer of 1000 nodes. For each model the output layer either consisted of a SoftMax activation function followed by a 16 node output layer for Transition task T\textsubscript{t-soft} or a Sigmoid activation function followed by a single node output layer for Transition task T\textsubscript{t-hard}.

We further configured our models by freezing the shallower half of all convolutional layers. Freezing layers is a common strategy performed when fine-tuning models for two main reasons: the first is to mitigate overfitting by reducing the total number of trainable parameters in the network; the second is to avoid redundant learning of low-level generic features that are already learned in our Source models \cite{pan2009survey}. 

After our models were configured for the Transition step. We split the Transition data into 80:20 Train and Validation sets, respectively. We then fine-tuned each model with Stochastic Gradient Descent and a batch size of 32, saving the weights that yielded the highest validation accuracy. Sparse-Categorical Cross Entropy loss was utilized for the Transition task T\textsubscript{t-soft} and Binary Cross Entropy loss was utilized for the Transition task T\textsubscript{t-hard}.

\textbf{Target step:} To begin the Target step we once again randomly initialized the pooling and fully connected layers of each model. This re-initialization ensures that we are fine-tuning the fully connected layers only on the Target task T\textsubscript{targ}. The binary nature of T\textsubscript{targ} moreover required a replacement of the SoftMax activation function with a Sigmoid activation function in soft labels models. We also added a dropout layer of 0.5 prior to the output layer of each model, to further mitigate overfitting. 

\begin{table}[h]
\centering
\includegraphics[scale=0.5,width=\linewidth]{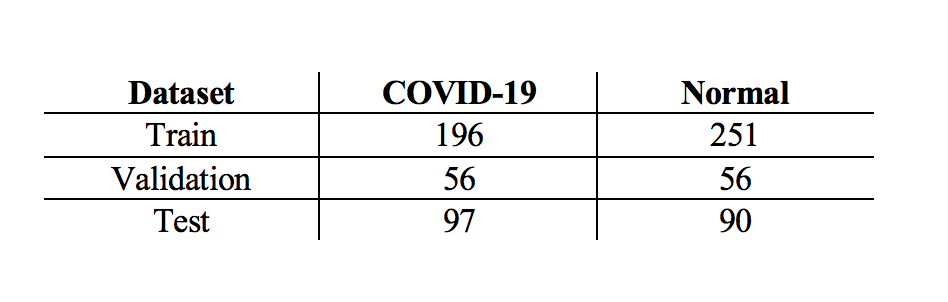}
\label{table:Table2}
\caption{Division of Target dataset}
\end{table}
As shown in Table 2, the Target dataset was split into 60:15:25 Train, Validation and Test sets, respectively. We then fine-tuned each model with Stochastic Gradient Descent and a momentum of 0.9 for 60 epochs and a batch size of 32, saving the weights that yielded the highest validation accuracy.
\vspace{2.5mm}

\section{Results}
In this section we present the results of our experiment outlined in the Methods section. As this paper aims to enhance the process of traditional Transfer Learning with a multi-source process, we compare the performances of our MSTL models against that of their baseline models pretrained on the ILSVRC dataset and fine-tuned on the Target dataset, referred to as the 'ImageNet' models. The performances of our finalized models are compared against their baselines to assess the magnitude of positive or negative transfer. 

After considering our two alternate labeling strategies used on our four models, we are left with 8 finalized MSTL models: ResNet50V2: Soft Labels, ResNet101V2: Soft Labels, DenseNet121: Soft Labels, DenseNet169: Soft Labels, ResNet50V2: Hard Labels, ResNet101V2: Hard Labels, DenseNet121: Hard Labels, and DenseNet169: Hard Labels. 

\begin{figure*}[h]
\centering
\includegraphics[scale=0.425]{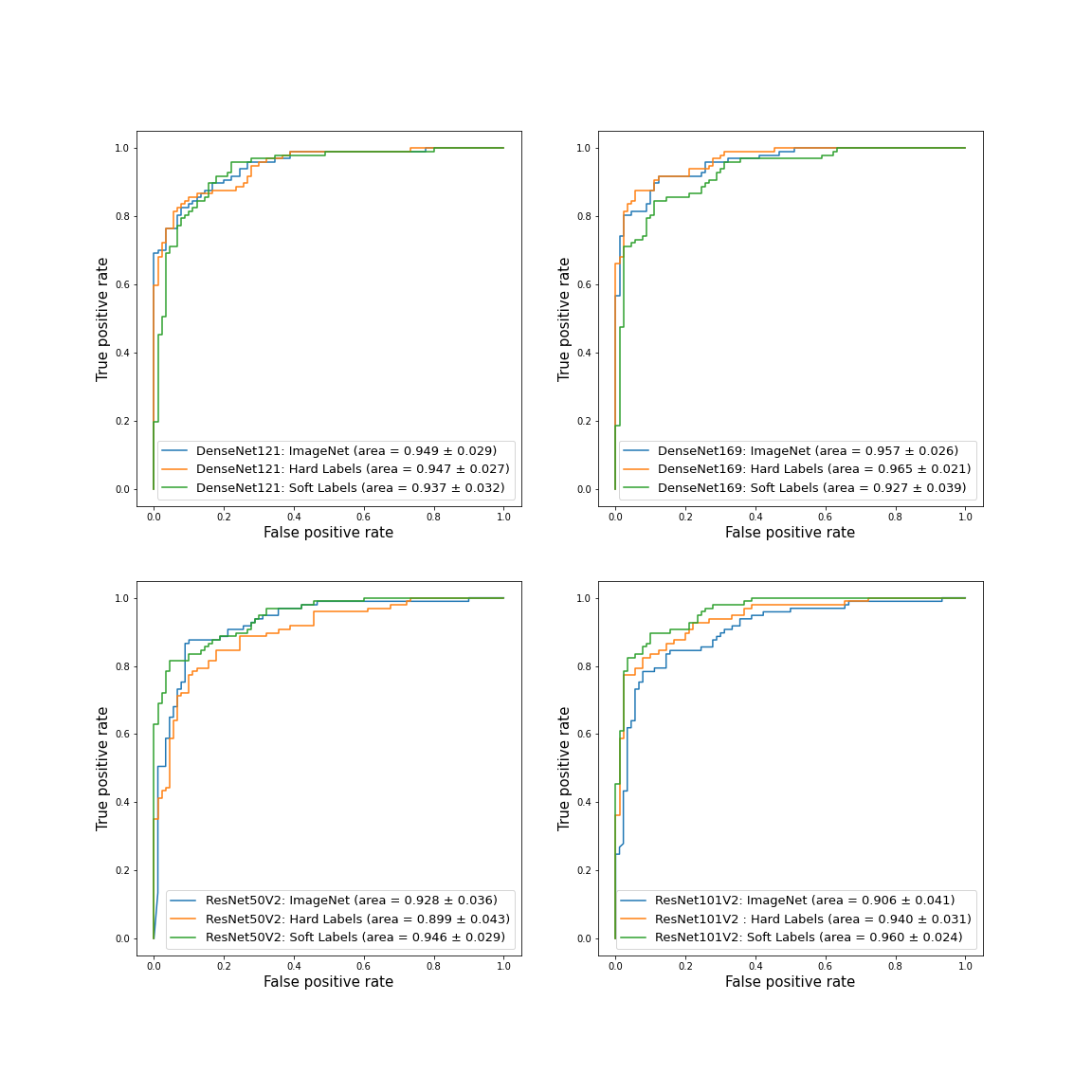}
\label{fig:Fig5}
\caption{ROC Curve per model architecture. True positive rate is plotted on y-axis, while false positive rate is plotted on x-axis.}
\end{figure*}

We first plotted the Receiver Operating Characteristic (ROC) of our finalized models against their baseline 'ImageNet' models isolated by model architecture. The ROC curve plots true positive rate against the false positive rate of samples predicted from the Test set. As shown in Figure 5, when using a Hard Labeling strategy both ResNet50V2 and ResNet101V2 models outperformed their baseline AUCs by 1.8\% and 5.4\%, respectively. However, their performances were not as exceptional when utilizing a Soft Labeling strategy, as only the ResNet101V2 outperformed its baseline AUC by 3.4\%, while the ResNet50V2 underperformed by 2.9\%. For the DenseNets, when using a Hard Labeling strategy, the deeper DenseNet169 outperformed its baseline AUC by 0.8\%, however, the shallower DenseNet121 underperformed by 0.2\%. When using a Soft Labeling strategy, both DenseNet121 and DenseNet169 underperformed by 1.2\% and 3.0\%, respectively.

\begin{figure*}[h]
\centering

\includegraphics[scale=0.45]{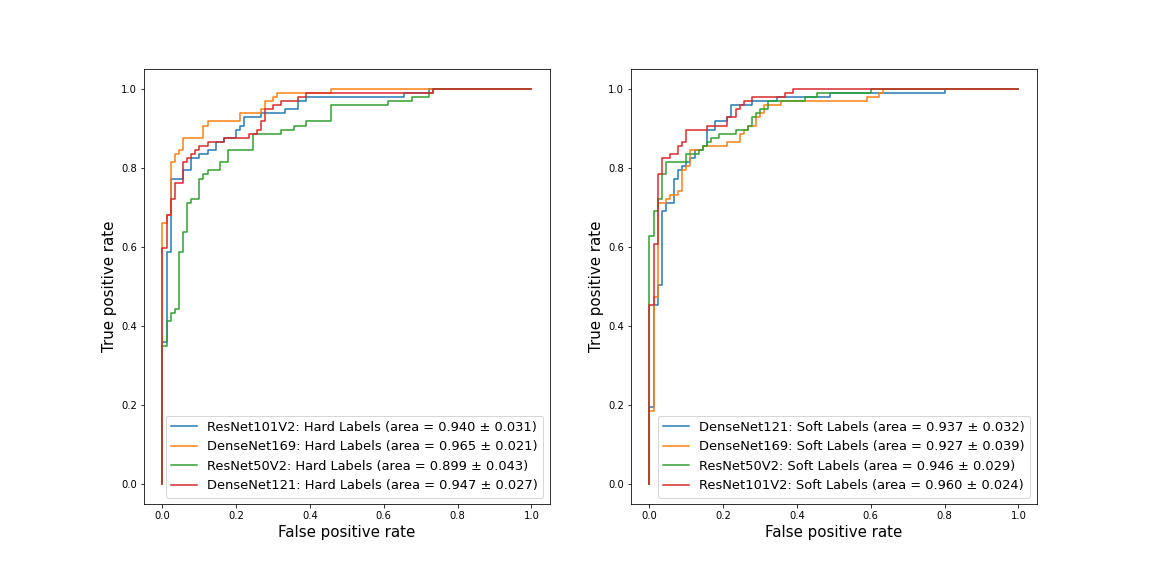}
\label{fig:Fig6}
\caption{ROC Curve per labeling method. True positive rate is plotted on y-axis, while false positive rate is plotted on x-axis.}
\end{figure*}

We then plotted the ROC of our finalized models isolated by labeling strategy. As shown in Figure 6, our DenseNet models outperformed our ResNet models by a margin of at least 0.7\% when employing a Hard Labeling strategy, and our ResNet models outperformed our DenseNet models by a margin of at least 0.9\% when employing a Soft Labeling strategy. When using Hard Labels the DenseNet169 model achieved a superior AUC of 0.965, followed by the DenseNet121 with an AUC of 0.947. When using Soft Labels the ResNet101V2 model achieved a superior AUC of 0.960, followed by the ResNet50V2 with an AUC of 0.946.

\begin{table*}

\centering
\includegraphics[width=12cm,height=9cm]{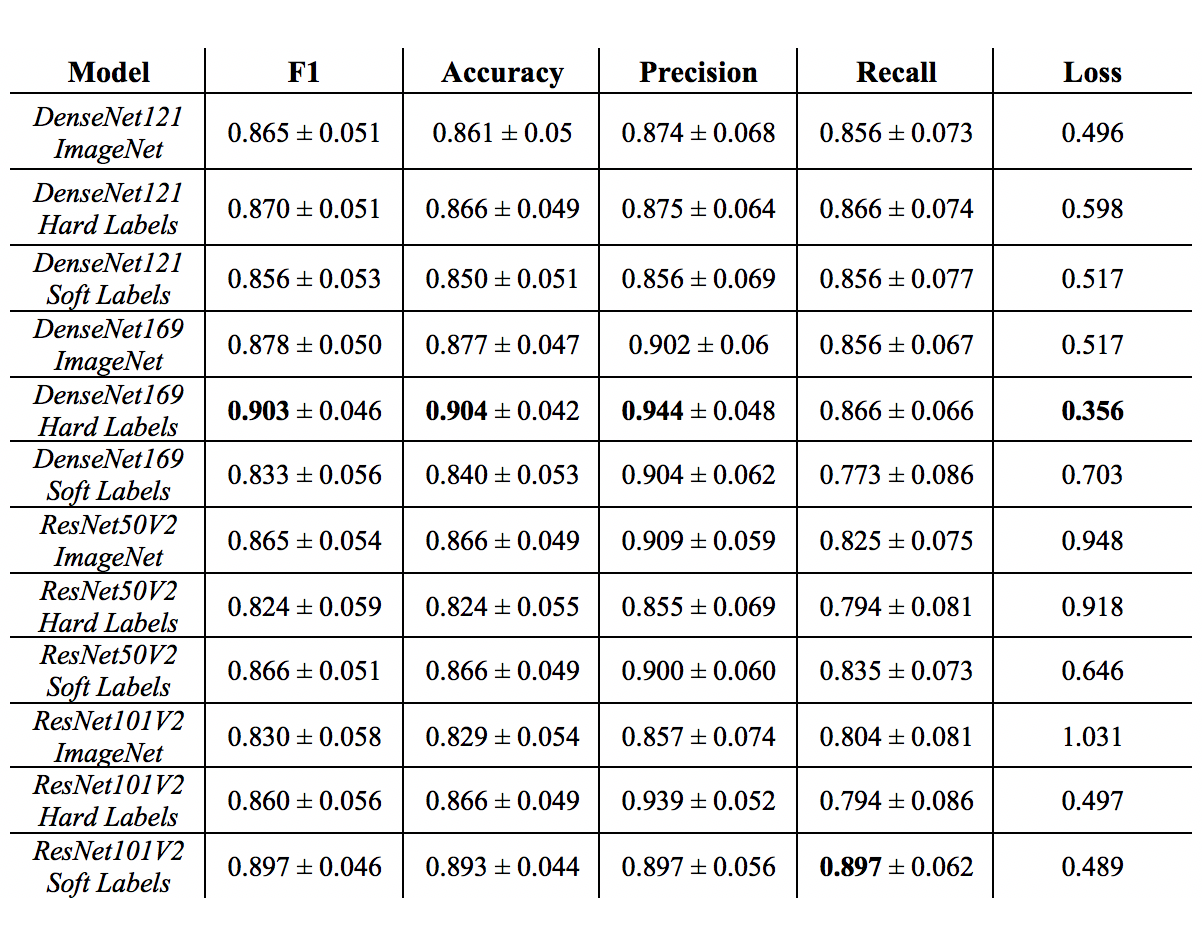}
\caption{F1, Accuracy, Precision and Recall scores}
\label{TABLE III}

\end{table*}
F1, Accuracy, Precision and Recall scores were then calculated between the models to further evaluate the models performances. As depicted in Table 3, the DenseNet169: Hard Labels model achieves superior F1, Accuracy, and Precision scores of 0.903, 0.904, and 0.944 while the ResNet101V2: Soft Labels model achieves a superior Recall score of 0.897. 
\vspace{2.5mm}

\section{Discussion}
When evaluating the performances of our models it is essential that we are reminded of our research objective. As stated in the Introduction section, our purpose in this study is to develop a highly sensitive classification model for the detection of COVID-19. We seek to emphasize the use of 'sensitive', because the sensitivity of the model takes precedence over all other metrics in the case of a medical imaging diagnosis. In other words, the cost of a False Negative greatly surpasses the cost of a False Positive, especially in the diagnosis of an infectious disease. In the case of a False Positive, the worst outcome would be that a individual without COVID-19 is told to self-isolate for two weeks. In the case of a False Negative, the worst outcome would be that an individual with COVID-19 continues to spread the infection after receiving a negative diagnosis.\\
\begin{figure*}[h]
\centering
\includegraphics[scale=0.7]{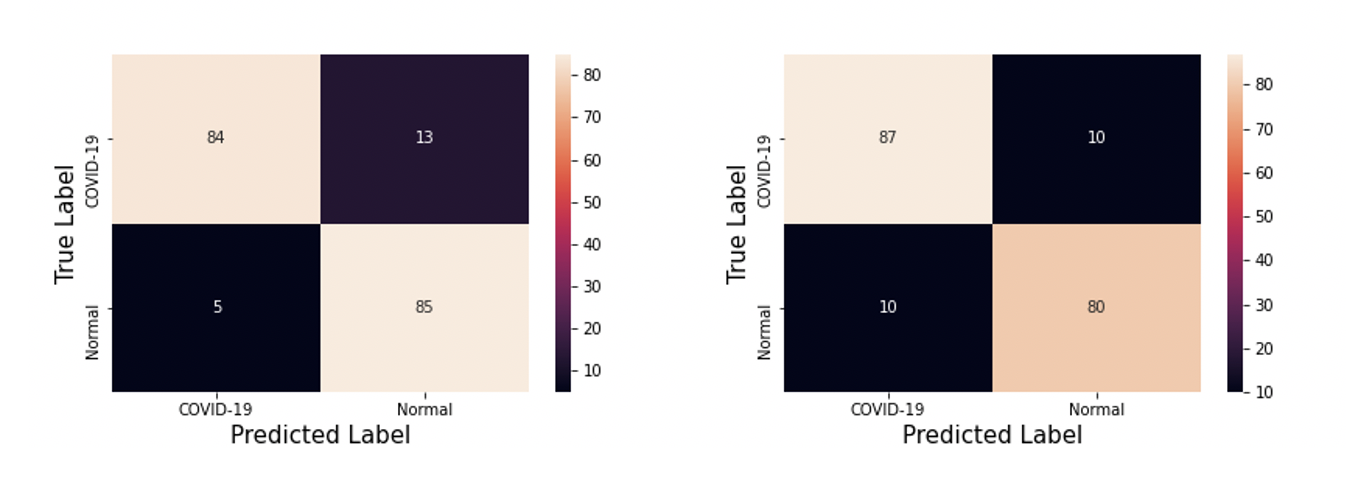}
\label{fig:Fig7}
\caption{DenseNet169: Hard Labels (left) \& ResNet101V2: Soft Labels (right)}
\end{figure*}

Therefore, the model with the highest Recall score is our most desirable model as it has the smallest chance of producing a False Negative. The classification results between our two highest performing models are visualized in Figure 7. As shown in the confusion matrices, although the DenseNet169: Hard Labels model has total fewer misclassifications than the ResNet101V2: Soft Labels model, the ResNet101V2: Soft Labels model has 3 fewer False Negative predictions, and thus is more desirable.

It should be noted that a significant finding of this study was the variance in performances between our DenseNets and ResNets with respect to the labeling strategy. As stated in the Results section, the DenseNets outperformed the ResNets when utilizing a Hard Labeling strategy, while the ResNets outperformed the DenseNets when utilizing a Soft Labeling strategy. We attribute these discrepancies in performance largely to the number of parameters in our models. As shown in Table 1, the ResNets are much more complex than DenseNets as they retain a larger number of total parameters. This greater complexity most likely caused the ResNets to overfit more to the binary classification Transition task T\textsubscript{t-hard} than to the multi-class classification task T\textsubscript{t-soft}. However, to conclude these hypotheses, further analysis is required to assess the relationship between the number of classes in T\textsubscript{t-soft} and the performances of our models.

\subsection{Limitations \& Future Work}
Although our results appropriately reflected the aspirations of our research objective, certain limitations apply to this study. 

The first limitation to this study is that CT scans are difficult to implement for mass testing. While scan results would be returned much quicker than RT-PCR Assays, CT scanning would need to be conducted indoors and under the same machine for thousands of patients. Due to the airborne nature of this infectious disease, an indoor testing environment is anything but ideal. 

The second limitation to this study is that CT scans are usually stored as 3-dimensional DICOM files, while our study requires an input of a 2-dimensional axial slice. This issue was exhibited in section 2.2, when our Transition dataset of 3-dimensional scans needed to be manually decomposed into relevant 2-dimensional slices. This preprocessing step was very computationally expensive, as it required manual selection of 10,176 CT slices displaying the regions of interest. To prevent this limitation, our study can be improved by automating the slice selection process. In this scenario, we could retain our original 2-dimensional architecture and Target dataset. The only expense would be training a independent classifier to assess if a given slice retains the region of interest we seek. 

A third limitation to this study is that confounding diseases can disrupt the performances of our models. As our models were trained to distinguish COVID-19 from normal CT scans, it runs the risk of classifying other lung diseases as False Positives (i.e. Influenza A). Although this presents an issue, we can diminish the risk of this type of missclassification by including other diseases in our Target dataset.
\vspace{2.5mm}

\section{Conclusion}
In this study we presented a Multi-Source Transfer Learning approach for the classification of COVID-19 from CT scans. By learning to classify an additional dataset of images more closely related to the Target domain, our models were able to outperform baseline models fine-tuned with traditional Transfer Learning methods. We additionally proposed an unsupervised label creation process, which further improved the performances of our Deep Residual Networks. The results of this study show the following: Transfer Learning can be improved by bridging the gap between the Source domain and the Target domain with a target-related Transition domain; unsupervised label creation has the potential to improve the performance of Deep Residual Networks; and with limited data, the application of Computer Vision for the detection of COVID-19 from CT scans exhibits high sensitivity and should be further investigated with the discussed limitations in mind.
\vspace{2.5mm}
\section{Acknowledgements}
This work was initiated and facilitated by CS 89.20/189 - Data Science for Health (Spring 2020) at Dartmouth College, taught by Professor Temiloluwa Prioleau.
\vspace{2.5mm}

\bibliographystyle{ieeetr}
\bibliography{references}

\end{document}